\documentclass[aps,prl,preprintnumbers,showpacs,nofootinbib]{revtex4}
\usepackage[utf8]{inputenc}
\usepackage[english]{babel}
\usepackage{amsmath}
\usepackage{amsfonts}
\usepackage{amssymb}
\usepackage{color}
\usepackage{slashed}
\usepackage{enumerate}
\usepackage{graphicx}
\usepackage{bm}
\usepackage{braket}
\usepackage{epstopdf}
\usepackage[usenames,dvipsnames,svgnames]{xcolor}
\usepackage[colorlinks=true,
            linkcolor=red,
            urlcolor=gray,
            citecolor=blue]{hyperref}

\providecommand{\NEG}{\slashed}
\begin{document}

\title{%
\textcolor{blue}{\textbf{Neutrino self-energy with new physics effects in an external magnetic field}}}
\author{Andr\'es Castillo$^{(a,*)}$}
\author{Rodolfo A. Diaz$^{(a)}$}
\author{Carlos G. Tarazona$^{(a,b,*)}$}
\author{John Morales$^{(a)}$}

\affiliation{$^{(a)}$ Universidad Nacional de Colombia, Sede Bogot\'a, Facultad de
Ciencias, Departamento de F\'{\i}sica. Ciudad Universitaria 111321,
Bogot\'a, Colombia}
\affiliation{$^{(b)}$Departamento de Ciencias B\'asicas. Universidad Manuela Beltr\'an.
Bogot\'a, Colombia}
\affiliation{$^{(*)}$ On leave at Fermilab-Theory Division, Pine Street $\&$ Kirk Road, Batavia, IL 60510}
\date{\today}

\begin{abstract}
We compute the magnetic dipole moment (MDM) for massive flavor neutrinos using the neutrino self-energy in a magnetized media. The framework to incorporate neutrino masses is one minimal extension of the Standard Model in which neutrinos are Dirac particles and their masses coming from tiny Yukawa couplings from a second Higgs doublet with a small vacuum expectation value. The computations are 
carried out by using proper time formalism in the weak field approximation $eB<<m_{e}^{2}$ and assuming normal hierarchy for neutrino masses and sweeping the charged 
Higgs mass. For $\nu_{\tau}$, analyses in the neutrino specific scenario indicate magnetic dipole moments greater than the values obtained to the MDM 
in the SM (with and without magnetic fields) and other flavor conserving models. This fact leading a higher proximity with experimental bounds and so on it is possible to get stronger exclusion limits over new physics parameter space.
\end{abstract}

\maketitle



\section{Introduction}
Electromagnetic properties of neutrinos have recently attracted much attention since they can be used to address many open questions in neutrino physics. From the theoretical point of view, electromagnetic form factors could contain information on new physics effects, which additionally solve other intriguing phenomena. One example is the oscillation mechanism and hierarchic structure of fermions, both explained with massive neutrinos and additional feasible beyond standard model fields. Another benchmark for electromagnetic behavior of neutrinos come from experimental and observational (astrophysical) effects that can be studied in several factories and detectors for neutrinos of all sources (atmospheric, solar, cosmological, and geoneutrinos)\cite{Giunti2015}. By analyzing these phenomena is possible to extract features over dynamically relevant interactions with detectors. In those observational scenarios, neutrinos and other particles can interact with external magnetic fields; which considerably affect the dynamics and electromagnetic properties. Also, the study of generation, propagation, and diffusion in 
 media of neutrinos in a magnetic field is important in several astrophysical contexts \cite{Raffelt}
 (e.g. in supernovae explosions, neutron stars formation, magnetars, and pulsars) and in early cosmology (e.g. 
 in the production of primordial magnetic fields and high neutrino density). 
  In all these cases the scale of magnetic field strength is greater than 
 $B_{e}=m_{e}^{2}/e\approx 4.41\times 10^{13}$ G where $m_{e}$ is the electron mass and $e$ 
 the elementary charge.That is a critical scale where magnetic fields have a strong impact on quantum processes as is shown by the Schwinger treatment of particle theories with external electromagnetic fields\footnote{The Schwinger mechanism described in the proper time formalism is a non-perturbative phenomenon by which particle-antiparticle pairs are produced by a static external field. In this frame, a tunneling process takes place between negative and positive energy states induced by the external potential. Due to the tunneling probability is exponential in $1/B$, probability of particle production is of order one only for fields that are comparable to a critical field $B\sim B_{e}=m_{e}^{2}/e$ \cite{Gelis}.}

Our aim is searching the link between theoretical and phenomenological consequences of electromagnetic properties for neutrinos, through the determination of the relationship between form factors as magnetic dipole moments and well motivated new physics particles. To describe such related processes is necessary to extend the traditional neutrino quantum field theory with a new regimen involving magnetic fields affecting manifest covariance of the former scenario.  These effects of magnetic fields on the electromagnetic properties and quantum fields of neutrino have been studied comprehensively in \cite{zhukovsky,Ioannisian}. In particular, a method to compute 
 the changes produced in the neutrino propagation is based on the modification of neutrino 
 self-energy operator by a media with a magnetic field \cite{Kuznetsov,Dobrynina,Dobrynina2}. When we take into account the media with a external magnetic field,   the neutrino electromagnetic form factors can be calculated from the self-energy using the proper-time 
(Schwinger) formalism  \cite{Erdas,Kuznetsov:2015uca}.

 In our study, we describe the initial effects on dispersion relations from new physics to neutrino MDM computing from self-energy method in the weak field approximation. The main goal is determining the additional contribution to the electromagnetic properties when a massive flavor active neutrino, emerging in a new physics scenario with charged Higgs bosons ($H^{\pm}$), goes into a magnetized media. To that end, this paper is organized as follows: In section \ref{sec:I} we review MDM contribution in the SM framework. A survey for our computations for 2HDM-$\nu$ fundamentals is given in section \ref{sec:II}. Also in section \ref{sec:III}, we present the extrapolation to obtain the MDM of new physics fields from 2HDM-$\nu$. Discussion over parameter space constrained by experimental data is performed in section \ref{sec:IV}.

\section{Magnetic dipole moments in the proper time formalism}
\label{sec:I}

In this section, we review the general structure for MDM with an external magnetic field by using effective Lagrangian approach and propagators obtained from proper time formalism. 

\begin{figure}[htp]
\centering
\includegraphics[scale=0.2]{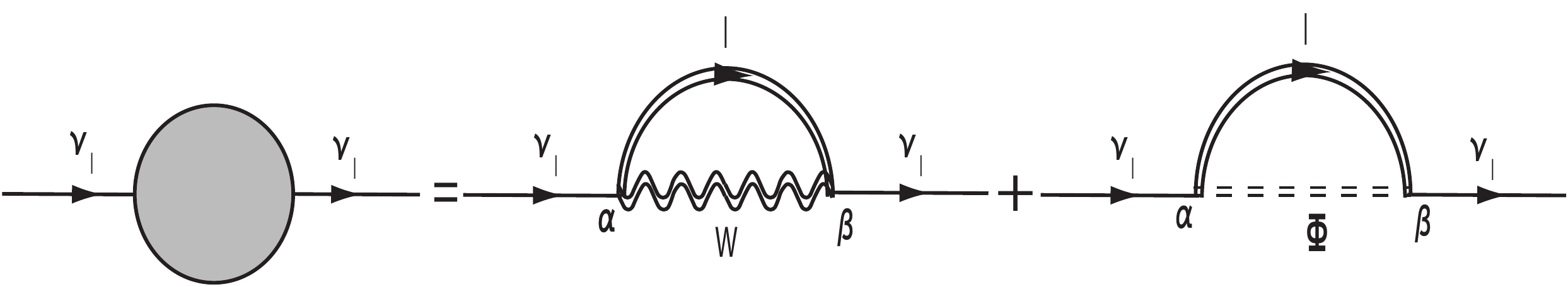} \vspace*{-0.4cm}
\caption{\textit{Feynman diagram describing the contribution to
self-energy operator for a flavor neutrino due to a constant and uniform magnetic field. The double line
corresponds to the charged lepton }$l$\emph{, the }$W^{\pm }$\emph{\ boson, the
charged Goldstone boson }$\Phi $ in an external magnetic field.}
\label{self_ener}
\end{figure}

In the effective approach, the change of neutrino energy in a magnetic field is due to a presence of a magnetic moment $\mu_{\nu}$  for the neutrino. The effective Lagrangian for the interaction, between neutrino field ($\psi$) and an external electromagnetic field, reads

\begin{align}
 \mathcal{L}^{(\mu)}=-i\frac{\mu_{\nu_{l}}^{B}}{2}\bar{\psi}\sigma_{\alpha\beta}\psi F^{\alpha \beta},
\end{align}

where $F_{\alpha\beta}=\partial_{\alpha}A_{\beta}-\partial_{\beta}A_{\alpha}$ is the electromagnetic field tensor and $\sigma_{\alpha\beta}=(\gamma_{\alpha}\gamma_{\beta}-\gamma_{\beta}\gamma_{\alpha})/2$. Getting an explicit form of MDM demands of self energy computation for neutrino and time proper formalism forms to find altereted propagator structures by the magnetic field. We start by considering  Feynman diagrams contributing to the neutrino self-energy in SM, which are depicted
in Fig. \ref{self_ener}.

In the weak field approximation $eB<< m_{e}^{2}$, the new structure to the gauge bosons $G_{B}\left( p\right) $, Goldstone
bosons $D\left( p\right) $, charged Higgs $D_{H}\left( p\right) $ and
fermions $S_{B}^{F}\left( p\right) $ propagators due to the presence of an
external magnetic field are respectively described by 
\begin{align}
G_{B}\left( p\right) &=-\frac{ig_{\alpha \beta }}{p^{2}-m_{W}^{2}}-\frac{%
2\beta \varphi }{\left( p^{2}-m_{W}^{2}\right) ^{2}}+\mathcal{O}\left( \beta
^{2}\right) ,\\
\ D\left( p\right)& =\frac{i}{p^{2}-m_{W}^{2}}+\mathcal{O}%
\left( \beta ^{2}\right) ,  \\
D_{H}\left( p\right) &=\frac{i}{p^{2}-m_{H^{\pm }}^{2}}+\mathcal{O}\left(
\beta ^{2}\right) ,\\
 S_{B}^{F}\left( p\right) &=\frac{i\left( m-{\slashed{p}}%
\right) }{p^{2}-m^{2}}+\beta \frac{\left( m-{\slashed{p}}_{\parallel
}\right) }{2\left( p^{2}-m^{2}\right) ^{2}}\left( \gamma \varphi \gamma
\right) +\mathcal{O}\left( \beta ^{2}\right),  \label{propagadores}
\end{align}
where $\beta =eB$ and $\ \varphi ^{\alpha \eta }=F^{\alpha \eta }/B$ is the
dimensionless electromagnetic field tensor normalized to $B$-field, with the
Lorentz indices of tensors are contracted as $\gamma \varphi \gamma =\gamma
_{\alpha }\varphi ^{\alpha \beta }\gamma _{\beta }$ and dual tensor $\tilde{%
\varphi}^{\alpha \eta }=\frac{1}{2}\epsilon ^{\alpha \eta \zeta \vartheta
}\varphi _{\zeta \vartheta }$. The decomposition of a four-vector in $\parallel $ (parallel component) and $\perp $ (perpendicular component) can be
defined by $p^{\mu }=p_{\parallel }^{\mu }+p_{\perp }^{\mu }=\left(
p^{0},0,0,p^{3}\right) +\left( 0,p^{1},p^{2},0\right) $, where $p_{\parallel }$ is the spatial part of momentum which is
parallel to the external $B$-field.

In its general form, the self-energy operator in an external $B$-field has the following Lorentz structure \cite{Erdas} 

\begin{equation}
\sum \left( p\right) =\left[ \mathcal{A}_{L}\slashed{p}+\mathcal{B}_{L}\slashed{p}_{\parallel
}+\mathcal{C}_{L}\left( p\tilde{\varphi}\gamma \right) \right] P_{L}+\left[ \mathcal{A}_{R}%
\slashed{p}+\mathcal{B}_{R}\slashed{p}_{\parallel }+\mathcal{C}_{R}\left( p\tilde{\varphi}\gamma
\right) \right] P_{R}+m_{\nu}\left[ \mathcal{K}_{1}+i\mathcal{K}_{2}\left( \gamma \varphi \gamma
\right) \right].  
\label{MDM_self}
\end{equation}%
The coefficients $A_{R},B_{R},B_{R}$ and $\mathcal{K}_{1,2}$ only come from the
Feynman diagram with the scalar $\Phi$ of Fig. \ref{self_ener}. The coefficients left, i.e., $\mathcal{A}_{L},\mathcal{B}_{L},\mathcal{C}_{L}$ are due to both Feynman diagrams in Fig. \ref{self_ener}. The coefficients $A_L$ , $A_R$ , and $K_1$ can be absorbed by the neutrino wave function and mass renormalization. The $B$ and $C$ terms are crucial for neutrino dispersion but, to lowest order, the dispersion relation depends only on $B_{L}$. The two coefficients $B_{R}$ and $C_R$ are suppressed
by a factor of $\epsilon_{\nu}$ relative to $B_{L}$ and $C_{L}$ couplings. Factors $K_{2}$, $C_{L}$ and $C_{R}$ are needed to construct the contribution from self-energy expansion
to neutrino MDM \cite{Dobrynina}, i.e.,  
\begin{equation}
\mu _{\nu _{l}}^{B}=\frac{m_{\nu }}{2B}\left( \mathcal{C}_{L}-\mathcal{C}_{R}+4\mathcal{K}_{2}\right).
\label{MDM_gen}
\end{equation}%
The result of this contribution calculated taking into account the charged $W
$-boson and charged scalar $\Phi $-boson is 
\begin{equation}
\mu _{\nu _{l}}^{B}=\mu _{\nu _{l}}\frac{1}{\left( 1-\lambda _{l}\right) ^{3}%
}\left( 1-\frac{7}{2}\lambda _{l}+3\lambda _{l}^{2}-\lambda _{l}^{2}\ln
\lambda _{l}-\frac{1}{2}\lambda _{l}^{3}\right),  
\label{MDM_result}
\end{equation}%
where $\mu _{\nu _{l}}=3.2\times 10^{-19}\mu _{B}\left( {m_{\nu _{\alpha }}}/{1 \text{
eV}}\right) $\ \ is the neutrino magnetic moment value in the vacuum \cite{Mohapatra,Giunti,broggini,Fukugita}
, $\mu _{B}={e}/{2m_{e}}$ is the Bohr magneton and  $\lambda _{l}={m_{l}^{2}}/{m_{W}^{2}}$.
 The value of the MDM in the vacuum and with the presence of a magnetic field for SM are reported in Tab. \ref{tab:EMNC}.

\begin{table}[tbph]
\begin{center}
\begin{tabular}{|l|l|l|l|}
\hline
& $m_{\nu }\left[ eV\right] $ & $\mu _{\nu }\left[ \mu _{B}\right] $ & $\mu
_{\nu }^{B}\left[ \mu _{B}\right] $ \\ \hline
$\nu _{e}$ & $0.06089$ & $1.948\times 10^{-20}$ & $1.948\times 10^{-20}$ \\ 
\hline
$\nu _{\mu }$ & $0.06754$ & $2.161\times 10^{-20}$ & $2.161\times 10^{-20}$
\\ \hline
$\nu _{\tau }$ & $0.07147$ & $2.287\times 10^{-20}$ & $2.286\times 10^{-20}$
\\ \hline
\end{tabular}%
%
%
\end{center}
\par
\vspace{-0.6cm}
\caption{\textit{Possibles values to MDM of the neutrinos in the vacuum and with magnetic
fields for SM. Flavor effective masses are values compatible with central values for PMNS
matrix \protect\cite{Forero} and with cosmological bounds \protect\cite%
{Abazajian,PAde} and differences over masses for $\protect\nu_{1},\protect\nu%
_{2}$ and $\protect\nu_{3}$ eigenstates in normal ordering \protect\cite{PDG}%
. The effective values for masses, obtained in this way, satisfy the bounds
for the average masses of the flavor eigenstates $m_{\protect\nu_{e}},m_{%
\protect\nu _{\protect\mu }},m_{\protect\nu _{\protect\tau }}<2.5$ eV 
\protect\cite{Fukugita}, and $m_{\bar{\protect\nu}_{e}}<2.05$ eV 
\protect\cite{PDG}.}}
\label{tab:EMNC}
\end{table}

\vspace{-0.6cm}

\section{Contribution to MDM in neutrino specific-2HDM\label{nspecific}%
}
\label{sec:II}
Before to introduce new physics effects in MDMs with magnetic fields, we make a survey for 2HDM-$\nu$
differentiating flavor properties and yielding mass mechanisms; which are relevant in the form of
interpreting electromagnetic properties for neutrinos.

The neutrino specific-2HDM (2HDM$-\nu $)  is implemented 
introducing a new scalar doublet $\Phi _{2}$ with the same quantum numbers as the SM Higgs doublet 
$\Phi _{1}$ and three right-handed neutrinos, which are singlets under the SM gauge group \cite{logan1,logan2}. 
The addition of the doublet $\Phi _{2}$ introduces five Higgs bosons in the scalar spectrum of neutrino specific structure. 
In general, 2HDM$-\nu $ is composed by two CP-even Higgs  $\left( H^{0},h^{0}\right) ,$ one CP-odd scalar $\left( A^{0}\right) 
$ and two charged Higgs bosons $\left( H^{\pm }\right) $ \cite{HHaber}.
In this model, the tiny neutrino masses could arise for small vacuum expectation value (VEV) of the doublet $\Phi _{2}$, 
which implies small masses or Yukawa couplings directly. 

The most general Lagrangian in 2HDM$-\nu $ is %
\begin{equation}
\mathcal{L}=\left( \mathcal{D}_{\mu }\Phi _{1}\right) ^{\dagger }\left( 
\mathcal{D}^{\mu }\Phi _{1}\right) +\left( \mathcal{D}_{\mu }\Phi
_{2}\right) ^{\dagger }\left( \mathcal{D}^{\mu }\Phi _{2}\right) -V_{H}+%
\mathcal{L}_{Y}.  \label{neutrinospecific}
\end{equation}

$V_{H}$ denotes the Higgs potential and which has the following form with a softly broken $U\left( 1\right) $ symmetry,

\begin{align}
V_{H}& =\bar{m}_{11}^{2}\Phi _{1}^{\dagger }\Phi _{1}+\bar{m}_{22}\Phi
_{2}^{\dagger }\Phi _{2}-\left( \bar{m}_{12}^{2}\Phi _{1}^{\dagger }\Phi
_{2}+h.c.\right)  \notag \\
& +\frac{1}{2}\lambda _{1}\left( \Phi _{1}^{\dagger }\Phi _{1}\right) ^{2}+%
\frac{1}{2}\lambda _{2}\left( \Phi _{2}^{\dagger }\Phi _{2}\right)
^{2}+\lambda _{3}\left( \Phi _{1}^{\dagger }\Phi _{1}\right) \left( \Phi
_{2}^{\dagger }\Phi _{2}\right) +\lambda _{4}\left( \Phi _{1}^{\dagger }\Phi
_{2}\right) \left( \Phi _{2}^{\dagger }\Phi _{1}\right) .
\label{HiggsPotential1}
\end{align}

On the other hand, $\mathcal{L}_{Y}$ in Eq. (\ref{neutrinospecific}) is the Yukawa Lagrangian (primordial for our studies), which in 2HDM-$\nu$ scenario reads 

\begin{equation}
-\mathcal{L}_{Y}=\xi _{ij}^{\nu }\bar{L}_{Li}\tilde{\Phi}_{2}\nu _{Rj}+\eta
_{ij}^{E}\bar{L}_{Li}\Phi _{1}E_{Rj}+\eta _{ij}^{D}\bar{Q}_{Li}\Phi
_{1}D_{Rj}+\eta _{ij}^{U}\bar{Q}_{Li}\tilde{\Phi}_{1}U_{Rj}+h.c.
\label{NELagrangian}
\end{equation}

$\tilde{\Phi}_{1}=i\sigma _{2}\Phi _{1}$ is the conjugate of the
Higgs doublet. Fermion doublets are defined by $Q_{L}\equiv \left(
u_{L},d_{L}\right) ^{T}$ and $L_{L}\equiv \left( \nu _{L},e_{L}\right) ^{T}.$
Here $E_{R}\left( D_{R}\right) $ is referred to the three down type weak
isospin lepton(quark) singlets and $\nu _{R}\left( U_{R}\right) $ is
referred to the three up type weak isospin neutrino (quark) singlets. The
first part in (\ref{NELagrangian}) comes from right-handed neutrinos, while
the second part is SM like. 

The lepton sector of the Yukawa Lagrangian, relevant for our computations, is given by%
\begin{equation}
-\mathcal{L}_{Y\left( 2HDM-\nu \right) } =\frac{\sqrt{2}m_{\nu
_{i}}}{v_{2}}\bar{\nu}_{l}\left( U_{PMNS}P_{L}\right) lH^{+}+h.c.
\label{couplings_ab}
\end{equation}

\subsection{Effective MDM in 2HDM-$\nu$}
\label{sec:III}
Guiding on scalar contributions to self-energy in SM, we compute an explicit form for the MDM contribution for new physics introduced by addtonal scalar fields in 2HDM-$\nu$.
\begin{figure}[htp]
\centering
\includegraphics[scale=0.25]{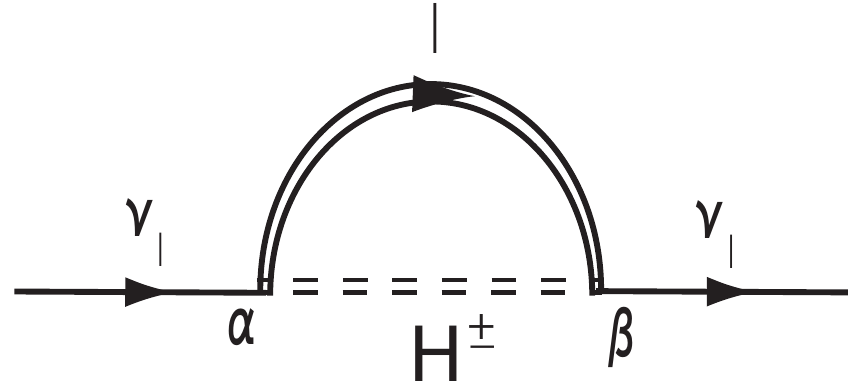} \vspace*{-0.4cm}
\caption{\textit{Feynman diagram describing the contribution to
self-energy operator for a neutrino in scenario with new physics due to a constant and uniform magnetic field. The double line
corresponds to charged lepton $l$ and charged Higgs boson $H^{\pm}$ in an external magnetic field.}}
\label{self_ener2}
\end{figure}

Within the framework of MDM in 2HDM$-\nu $, we should add one new type 
of contribution to self-energy operator of the neutrino due to charged Higgs bosons $H^{\pm }$ (see diagram in Fig. \ref{self_ener2}), i.e.,
\begin{equation}
\sum \left( p\right) =\sum\nolimits_{W^{\pm }}\left( p\right)
+\sum\nolimits_{\Phi }\left( p\right) +\sum\nolimits_{H^{\pm }}\left(
p\right).
\end{equation}%
The new physics effects separating the contributions of SM and 2HDM$-\nu $ coming from the 
contribution $\sum\nolimits_{H^{\pm }}\left(p\right)$ are
\begin{align}
\sum\nolimits_{H^{\pm }}\left( p\right) =i\int \frac{d^{4}k}{\left( 2\pi \right) ^{4}}\left( aP_{L}+bP_{R}\right)
S_{B}^{F}\left( p-k\right) \left( cP_{R}+dP_{L}\right) D_{B}\left( k\right) ,
\end{align}

where $S_{B}^{F}$ and $D_{B}$ are the propagators of the charged lepton and
Higgs charged boson, the constants $a,b,c$ and $d$ are associated with the Feynman rules of the
particular $\nu$-2HDM in the vertex, and, finally, $P_{L}=\left( 1+\gamma _{5}\right) /2$ and $%
P_{R}=\left( 1-\gamma _{5}\right) /2$  are the left-handed and right-handed chiral operators.  The new physics self-energy can be expanded by\footnote{The expansion shown in (\ref{self}) is model independent for the Yukawa couplings structure. Other models with charged Higgs bosons or charged scalars can be tested with this self-energy form by varying $a$ and $b$ couplings. These effects have been treated in \cite{Proceedings}.}

\begin{eqnarray}
\sum\nolimits_{H^{\pm }}\left( p\right)&=&-i\int \frac{d^{4}k}{\left( 2\pi \right) ^{4}}\left[ \frac{a^{2}\left( \NEG{p}-\NEG{k}\right) P_{R}}{\left( \left( p-k\right)
^{2}-m_{l}^{2}\right) \left( k^{2}-m_{H^{\pm }}^{2}\right) }-\frac{%
2a^{2}\beta \left( p\tilde{\varphi}\gamma \right) P_{R}-2a^{2}\beta \left( k%
\tilde{\varphi}\gamma \right) P_{R}}{\left( \left( p-k\right)
^{2}-m_{l}^{2}\right) ^{2}\left( k^{2}-m_{H^{\pm }}^{2}\right) }\right. \nonumber\\
&&\left.+\frac{%
abm_{l}P_{R}}{\left( \left( p-k\right) ^{2}-m_{l}^{2}\right) \left(
k^{2}-m_{H^{\pm }}^{2}\right) }-\frac{i\beta abm_{l}\left( \gamma \varphi
\gamma \right) }{\left( \left( p-k\right) ^{2}-m_{l}^{2}\right) ^{2}\left(
k^{2}-m_{H^{\pm }}^{2}\right) } \right. \label{self}\\
&&\left.+\frac{abm_{l}P_{L}}{\left( \left( p-k\right) ^{2}-m_{l}^{2}\right) \left(
k^{2}-m_{H^{\pm }}^{2}\right) }+\frac{b^{2}\left( \NEG{p}-\NEG{k}\right)
P_{L}}{\left( \left( p-k\right) ^{2}-m_{l}^{2}\right) \left( k^{2}-m_{H^{\pm
}}^{2}\right) }-\frac{2\beta b^{2}\left( p\tilde{\varphi}\gamma \right)
P_{L}-2\beta b^{2}\left( k\tilde{\varphi}\gamma \right) P_{L}}{\left( \left(
p-k\right) ^{2}-m_{l}^{2}\right) ^{2}\left( k^{2}-m_{H^{\pm }}^{2}\right) }%
\right], \notag
\end{eqnarray}

where $m_{l}$ and $m_{H^{\pm }}$ are the masses of the
charged lepton and $H$-boson respectively. Here coefficients  $a = c$ and $b = d$ in the vertices are equal in the Feynman rules for $\nu$-2HDM. Extrapolating scalar contribution from energy modifications with scalars of self energy due to effective MDM as in Eq. (\ref{MDM_self}), we factorize from (\ref{self}) new physics coefficients $\bar{\mathcal{C}}_{L}$, $\bar{\mathcal{C}}_{R}$ 
and $\bar{\mathcal{K}}_{2}$ belonging to the MDM contribution in the self-energy for charged Higgs boson	
\begin{eqnarray}
\bar{\mathcal{C}}_{L}\left( p\tilde{\varphi}\gamma \right) P_{L} &=&i\int \frac{d^{4}k}{%
\left( 2\pi \right) ^{4}}\frac{2\beta b^{2}\left( p\tilde{\varphi}\gamma
\right) P_{L}-2\beta b^{2}\left( k\tilde{\varphi}\gamma \right) P_{L}}{%
\left( \left( p-k\right) ^{2}-m_{l}^{2}\right) ^{2}\left( k^{2}-m_{H^{\pm
}}^{2}\right) },  \label{cl} \\
\bar{\mathcal{C}}_{R}\left( p\tilde{\varphi}\gamma \right) P_{R} &=&-i\int \frac{d^{4}k}{%
\left( 2\pi \right) ^{4}}\frac{2a^{2}\beta \left( p\tilde{\varphi}\gamma
\right) P_{R}-2a^{2}\beta \left( k\tilde{\varphi}\gamma \right) P_{R}}{%
\left( \left( p-k\right) ^{2}-m_{l}^{2}\right) ^{2}\left( k^{2}-m_{H^{\pm
}}^{2}\right) },  \label{cr} \\
m_{\nu }i\bar{\mathcal{K}}_{2}\left( \gamma \varphi \gamma \right)  &=&i\int \frac{d^{4}k}{%
\left( 2\pi \right) ^{4}}\frac{i\beta abm_{l}\left( \gamma \varphi \gamma
\right) }{\left( \left( p-k\right) ^{2}-m_{l}^{2}\right) ^{2}\left(
k^{2}-m_{H^{\pm }}^{2}\right) }.  \label{k2}
\end{eqnarray}%
Substituting the expressions (\ref{cl}),(\ref{cr}) and (\ref{k2}) into one MDM
new physics emulating (scalar) contribution of Eq. \ref{MDM_gen}, 
\begin{eqnarray}
\left( \mu _{\nu _{l}}^{B}\right) _{H^{\pm }} &=&\frac{m_{\nu }}{2B}\left(
\bar{\mathcal{C}}_{L}-\bar{\mathcal{C}}_{R}+4\bar{\mathcal{K}}_{2}\right)   \nonumber \\
&=&\frac{\sqrt{2}}{3G_{f}}\mu _{\nu _{l}}\int\limits_{0}^{1}dx\frac{2\left(
x^{2}-3x+2\right) \left( b^{2}+a^{2}\right) +\left( 1-x\right) \frac{m_{l}}{%
m_{\nu }}ab}{m_{\nu }^{2}x^{2}+\left( m_{l}^{2}-m_{\nu }^{2}-m_{H^{\pm
}}^{2}\right) x+m_{H^{\pm }}^{2}}.  \label{2HDM_B_MDM}
\end{eqnarray}%
where we have used
\[
\frac{m_{\nu }}{2B}\frac{\beta }{4\pi ^{2}}=\frac{\sqrt{2}}{3G_{f}}\mu _{\nu
_{l}},
\]%
the coefficients $a$ and $b$ coming from Eq. (\ref{couplings_ab}) are $a=-\frac{\sqrt{2}}{v_{2}}m_{\nu _{i}}U_{k,i}$
and $b=\frac{\sqrt{2}}{v_{2}}m_{\nu _{i}}U_{i,k}$. In our numerical analyses, we shall use the 
accurate value for Fermi's constant: 
$G_{F}=\protect\sqrt{2}g^{2}/8 M_{W}^{2}=1.1663787(6)\times 10^{-5}\hspace{0.1cm}\text{GeV}^{-2}$.

\section{Results and analysis for MDM in 2HDM$-\nu $\label{results}}
\label{sec:IV}

\begin{figure}[tph]
\centering
\includegraphics[scale=0.425]{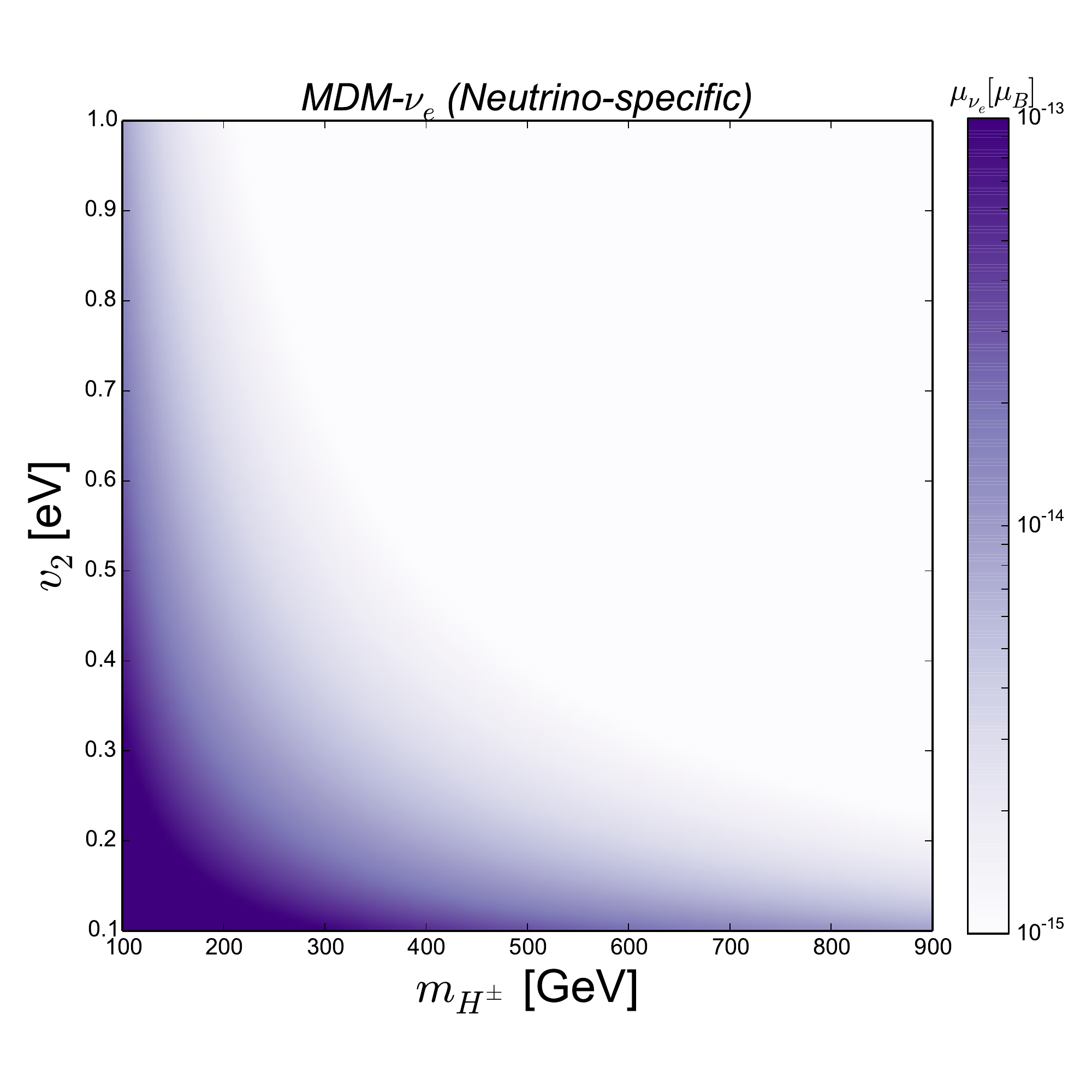}
\includegraphics[scale=0.425]{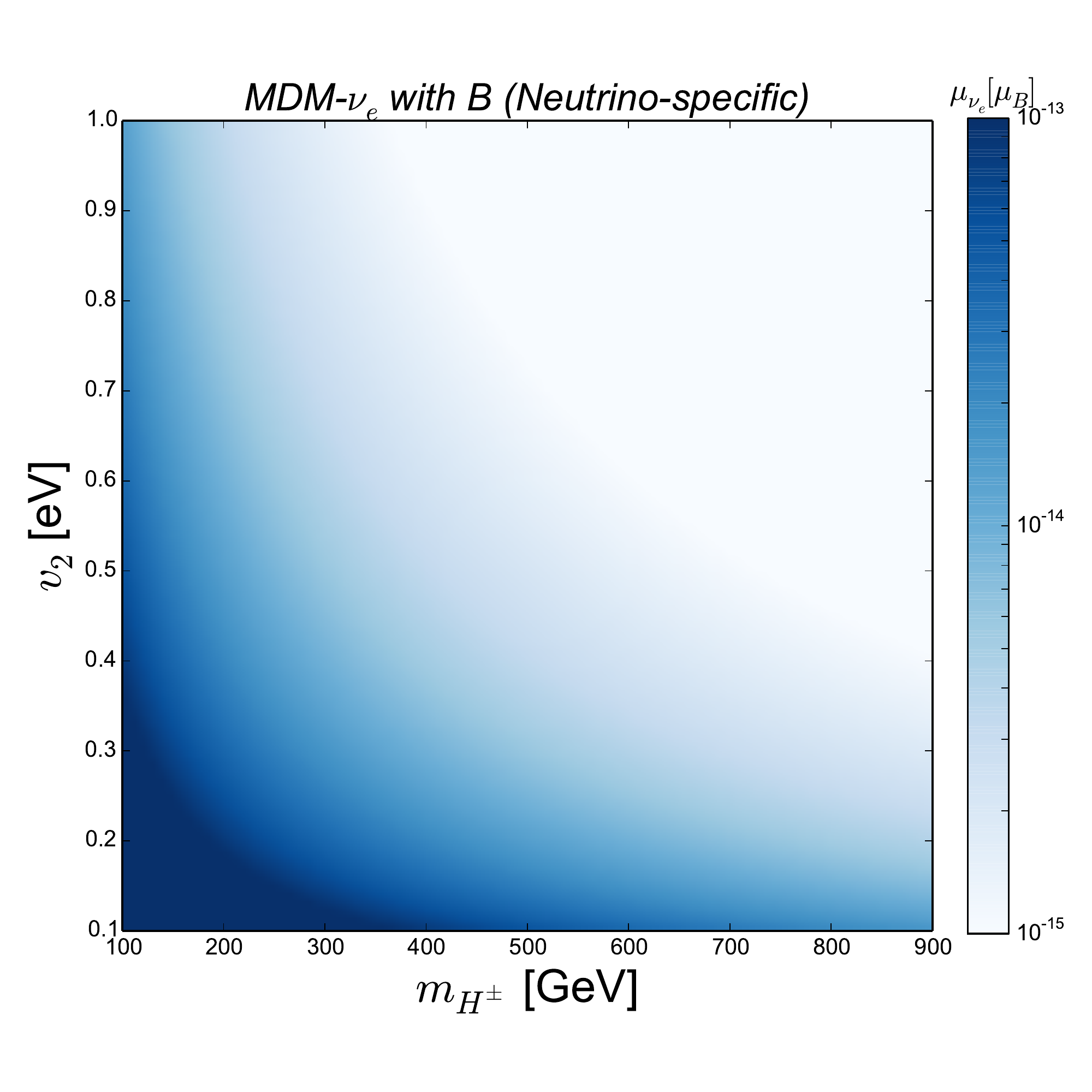}%
 \vspace{-0.6cm}
\caption{\textit{Contribution to the magnetic dipolar moment $\mu_{\nu_{e}}$  in vacuum (\textbf{Left}) and with presence of magnetic fields (\textbf{Right}) coming from the 2HDM$-\nu $ couplings. Analyses in vacuum are performed using traditional perturbative treatment in feynman diagrams with triangle vertex corrections \cite{IJMPA}. Numerical computations used effective neutrino masses of Tab. \ref{tab:EMNC}. }}
\label{specificI}
\end{figure}

\begin{figure}[tph]
\centering
\includegraphics[scale=0.425]{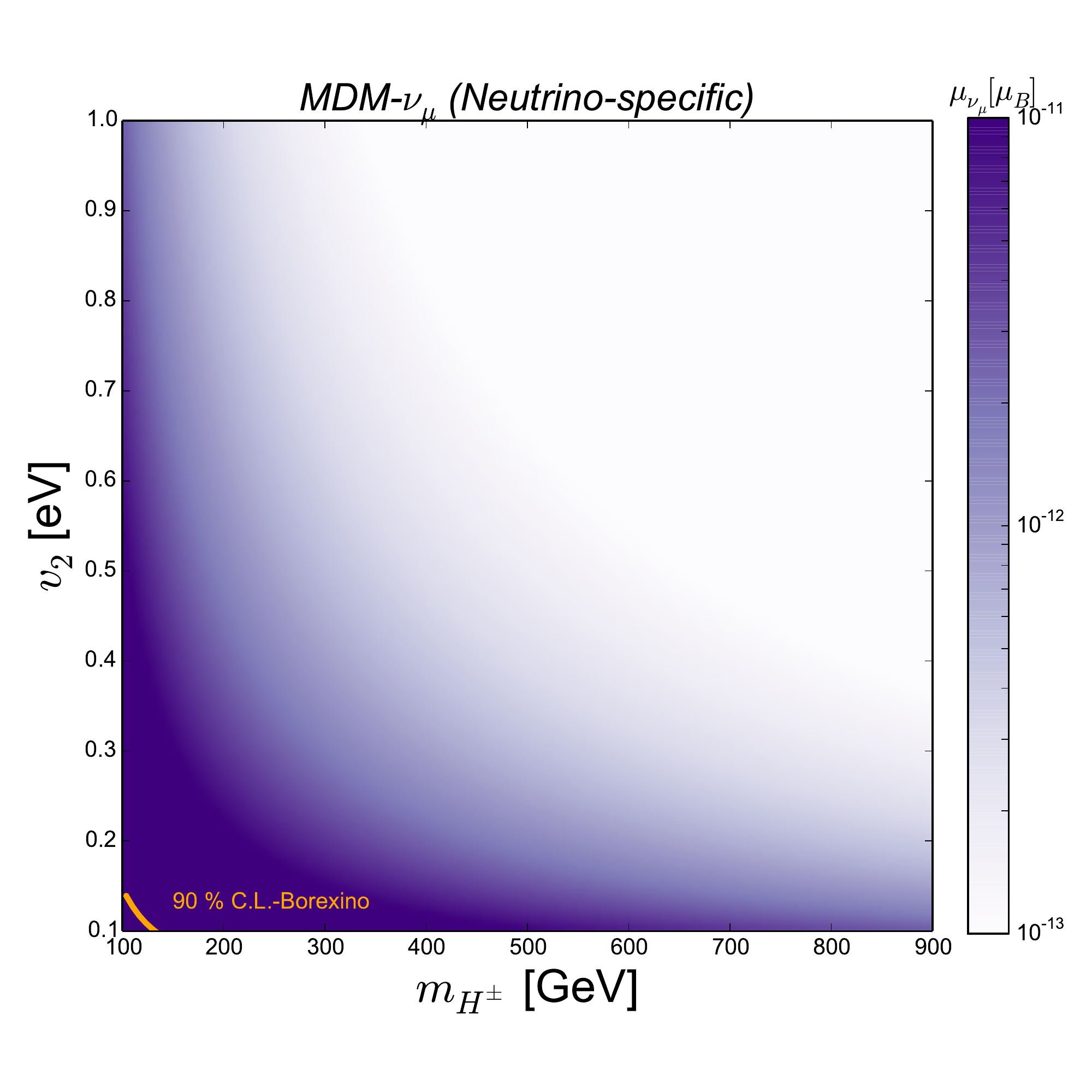}
\includegraphics[scale=0.425]{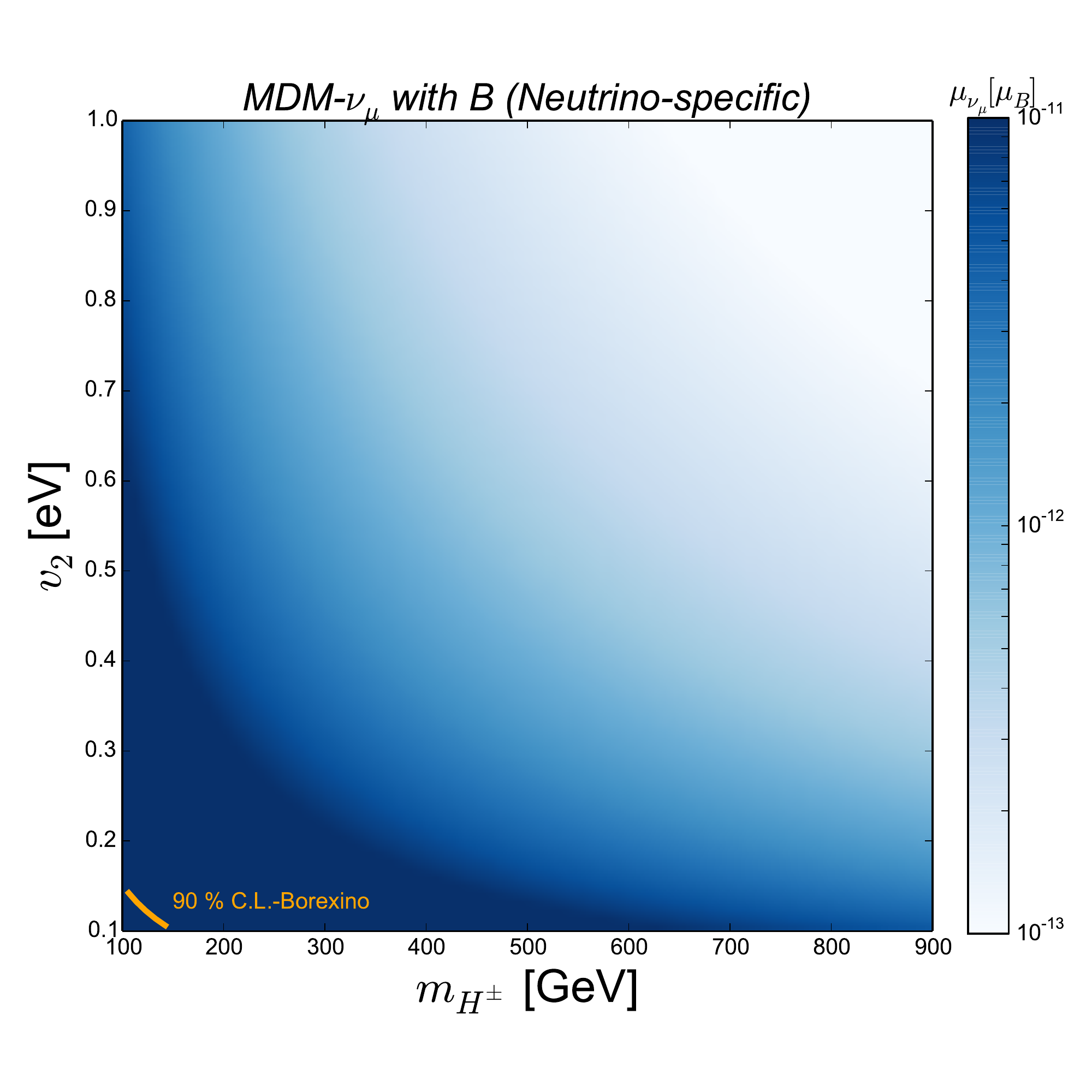}%
 \vspace{-0.6cm}
\caption{\textit{Contribution to the magnetic dipolar moment $\mu_{\nu_{\mu}}$ in vacuum (\textbf{Left}) and with presence of magnetic fields (\textbf{Right}) coming from the 2HDM$-\nu $ couplings. The contour line corresponds to the experimental upper limits for MDM
coming from BOREXino \cite{Montanino} that gives $\mu _{\nu _{\mu
}}<1.9\times 10^{-10}\mu _{B}$ at $90\%~$ C.L. Analyses in vacuum are performed using traditional perturbative treatment in feynman diagrams with triangle vertex corrections \cite{IJMPA}. Numerical computations used effective neutrino masses of Tab. \ref{tab:EMNC}.}}
\label{specificII}
\end{figure}

\begin{figure}[tph]
\centering
\par
\includegraphics[scale=0.425]{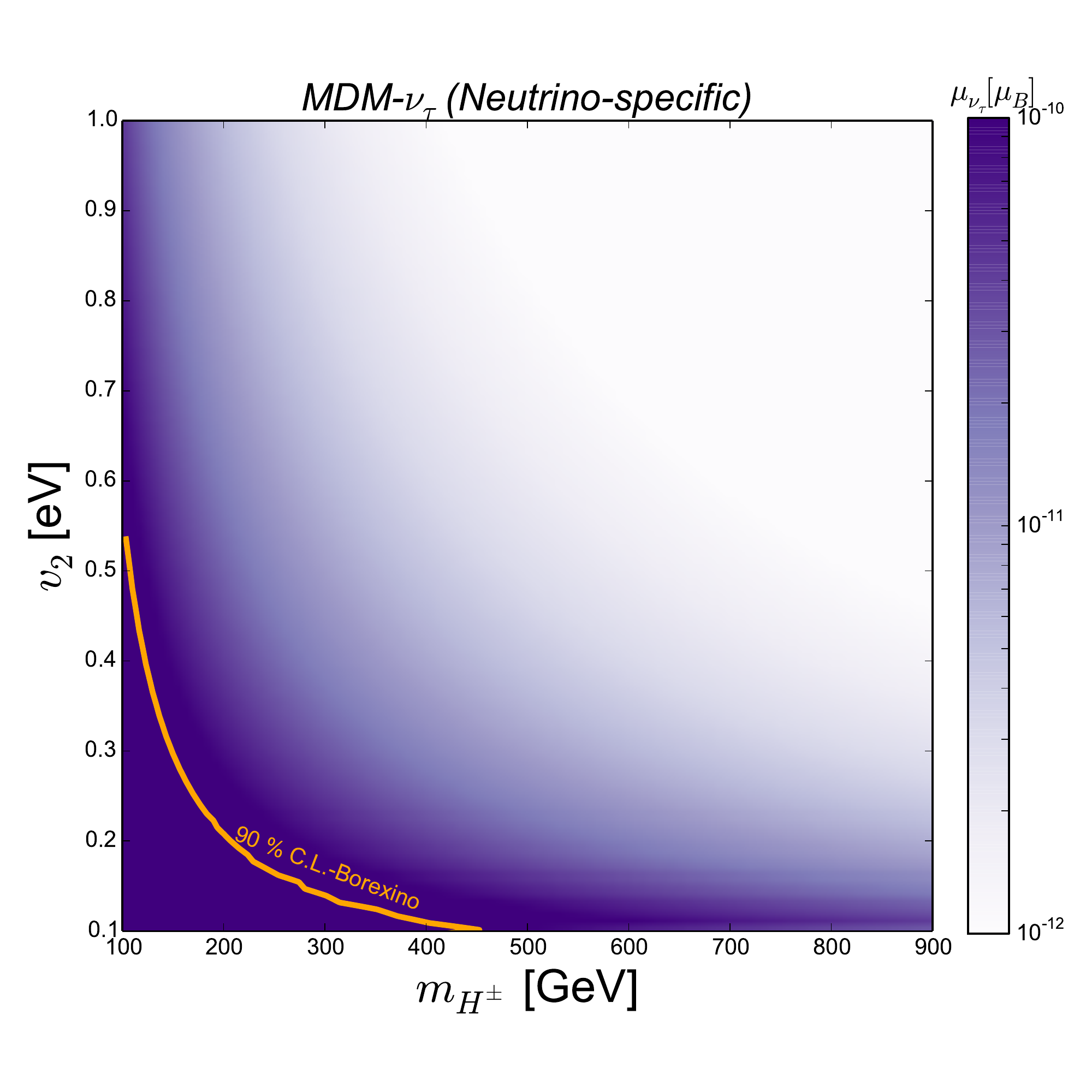}
\includegraphics[scale=0.425]{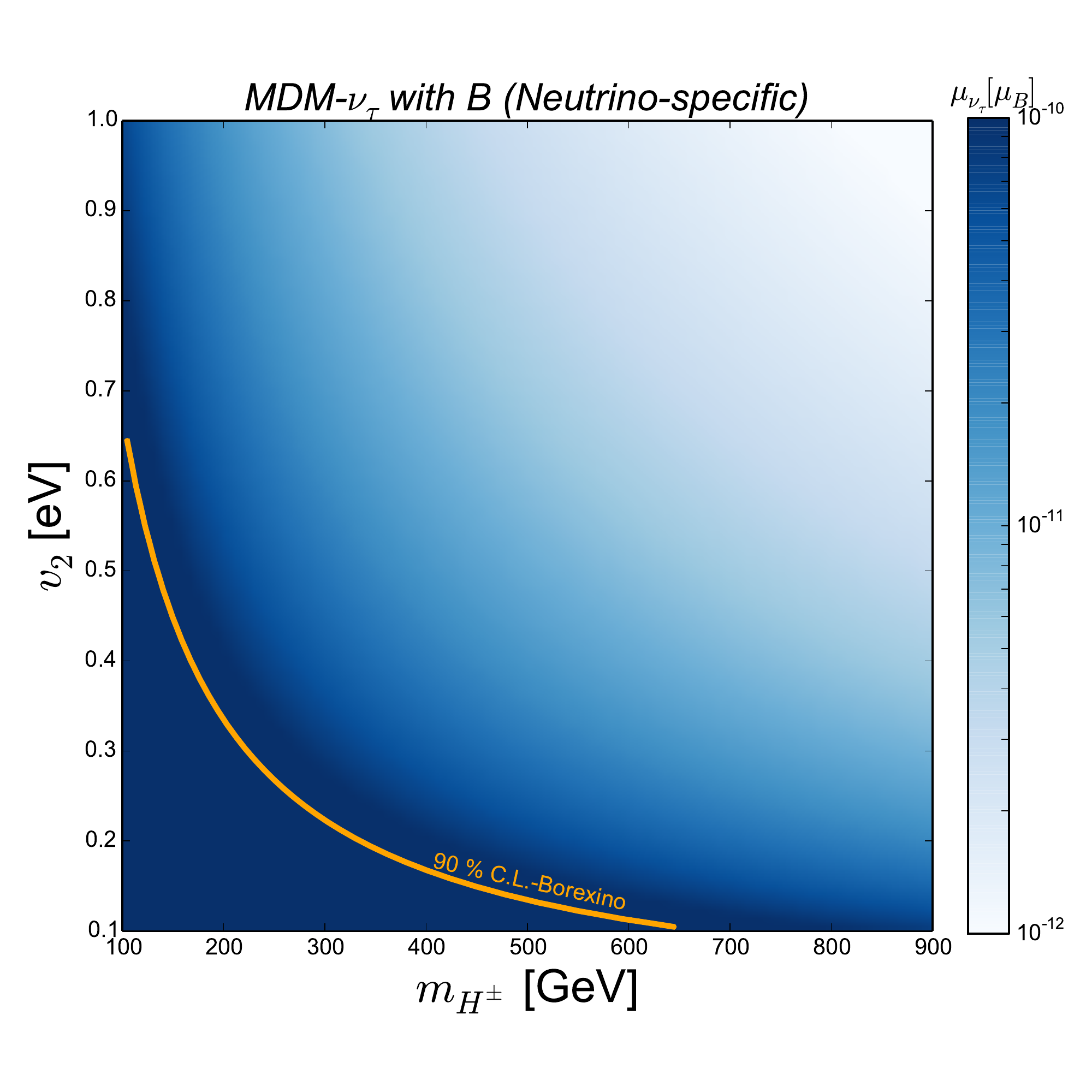} \vspace{-0.6cm}
\caption{\textit{Contribution to the magnetic dipolar moment $\mu_{\nu_{\tau}}$ in vacuum (\textbf{Left}) and with presence of magnetic fields (\textbf{Right}) coming from the 2HDM$-\nu $ couplings. Yellow lines makes
reference to the experimental thresholds for $\mu_{\nu_{\tau}}$ at 90$\%$ C.L of BOREXino
2008 \cite{Montanino}, whose upper limit is $\mu _{\nu _{\tau }}<1.5\times
10^{-10}\mu _{B}$ at $90\%~$ C.L. Analyses in vacuum are performed using traditional perturbative treatment in feynman diagrams with triangle vertex corrections \cite{IJMPA}.  Numerical computations used effective neutrino masses of Tab. \ref{tab:EMNC}.}}
\label{specificIII}
\end{figure}

Our analyses are based on constraints on charged Higgs masses and the VEV of the second doublet ($v_{2}$). The effect of the Yukawa structure of the Dirac neutrino in the 2HDM$-\nu $, which has the $m_{\nu }/v_{2}$ ratio, makes the MDM be strongly sensitive to the VEV and the type of hierarchy for neutrino masses ordering. To see those effects and parameter relevance in face of MDMs, we consider scannings (Figs. \ref{specificI}-\ref{specificIII}) showing the evolution of MDM for flavor neutrinos in the $(m_{H^{\pm}},v_{2})$ space.

In the case of electron neutrino (Fig. \ref{specificI}) and muon neutrino (Fig. \ref{specificII}), we see as the
contribution overpasses the SM-MDM ($\mu_{\nu}$) in several orders of magnitude (see Tab. \ref{tab:EMNC}), being only closed the 
experimental limits for values of charged Higgs masses close to $m_{H^{\pm }}=100$ GeV. Nevertheless, in these analyses it is not possible to get strong enough constraints over  $(m_{H^{\pm}},v_{2})$.

For $\nu_{\tau}$-neutrinos (see Fig. \ref{specificII}), we have one of the most
promissory scenarios, due to it overpasses the experimental thresholds leading constraints over space parameter of the 2HDM-$\nu$ scenario. However, as we approach near to the experimental
threshold, Yukawa couplings are also reaching the perturbativity limit.
Indeed, these limits are established to avoid divergences in the
energy evolution of renormalization group equations. The reason is that the value of Yukawa
coupling (square) cannot exceed the bound of $8\pi $ since beyond this limit
the chance to find out some Landau pole in energy couplings evolution is
significantly greater \cite{Kanemura}. This value translates into a lower
bound of $v_{2}$ for the respective neutrino mass present in the normal
hierarchy. In the case of $\nu _{\tau },$ for instance, $v_{2\text{min}%
}=0.015$ eV. Since the perturbation theory is reliable for values of $v_{2}>v_{2\text{min}}$, we choose $v_{2}=0.1$ GeV as the starting point of all scannings. We consider that lower
values of $v_{2}$ overpassing the threshold corrections are all
non-perturbative contributions to MDM and results can be strongly dependent on this starting value. Furthermore, in particular cases,
contributions to MDMs can be constrained by threshold experimental limits,
as for instance from $\nu_{\tau}$ contributions, the fact of $v_{2}>0.65$ eV
for values are allowed at 90 $\%$ C.L. for all charged Higgs masses.  At 90$\%$ C.L., values $v_{2}>0.1$(GeV) are allowed for charged Higgs masses $m_{H^{\pm}}>650$ GeV. By allowing smaller numbers of $v_{2}$ without to reach the perturbative limit,
the constraints also exclude several values of charged Higgs masses.

One sees as for 2HDM-$\nu$, corrections involving magnetic fields are higher than vacuum analyses. This result is model dependent, as can be established for other 2HDMs without feasible natural terms for neutrino masses. More specifically,  for 2HDMs type I, II and III contributions with external magnetic fields interfere destructively with vacuum contributions \cite{Proceedings}.  

\section{Concluding remarks\label{sec:conclusions}}

We have calculated the self-energy contribution to the neutrino MDM in a magnetized media using a non-minimal extension for SM with two Higgs doublets. Even though our computations are model independent, the most important contribution in 2HDMs studied so far comes from neutrino specific scenario. 
Also, the 2HDM$-\nu $ scenario could be a plausible scenario to search the MDMs by means of the intimate relation with neutrino mass and tunning of VEVs. It can be seen that all contributions of 2HDM-$\nu'$ scenario are above those obtained
from SM for all neutrino flavors. This due to the structure of the Yukawa couplings, 
the mass of the Higgs charged boson and charged lepton in the vertex,  which is the same flavor of the corresponding neutrino. Moreover, for this model, the introduction of magnetic effects increases the contribution to the neutrino MDM compared with the value of MDM obtained in vacuum.

In all effective flavor magnetic dipole moments, the largest contributions to the neutrino MDM are given for values tending to zero of the 
VEV of the second doublet, as well as for values close to $100$ GeV of the charged Higgs boson. In the case $\nu_{\tau}$, those values are constrained by Borexino experimental analyses at 90$\%$ C.L..

It is worthwhile to point out that the MDM effects may be slightly different due to changes in the hierarchy used for neutrino masses. Although it is important to 
emphasize that the mass of the charged lepton plays a relevant role in the 
 the effective MDM. These facts lead to a framework wherein the third family has the most significant values for MDMs.

From global analyses, $\nu _{\tau }$-MDM with magnetic fields could be converted into a plausible
reference point to study new physics scenarios contributing to electromagnetic
form factors. Particularly, in the light of models with feasible terms for neutrino masses. 
Avoiding regions where perturbativity could be in conflict with lower values of the natural
VEV $v_{2}$ and charged Higgs masses, the neutrino specific model would be a significant benchmark to
introduce these effects of new physics, at the same time that neutrino
masses are at one viable scale. This opens an additional window to study these electromagnetic effects in most elaborated models where the mass terms for neutrinos appears naturally via an appropriate see-saw mechanism. Our computations are relevant in those scenarios in the limit where additional gauge bosons and fermions become decoupled of the theory.

\section{Acknowledgments}

We acknowledge financial support from Colciencias and DIB (Universidad
Nacional de Colombia). Carlos G. Tarazona also thanks to Universidad Manuela
Beltran as well as the support from DIB-Project with Q-number \textbf{%
110165843163} (Universidad Nacional de Colombia). A. Castillo, J. Morales,
and R. Diaz are also indebted to the \emph{Programa Nacional Doctoral de
Colciencias} for its academic and financial aid. A. Castillo and Carlos. G. Tarazona are also indebted with the Summer Visitor Program given by Theory Department at Fermilab.

\end{document}